\documentstyle[aps,twocolumn]{revtex}
\begin{document}
\title{The Schrodinger Cat Family in Attractive Bose Gases and Their
Interference}
\author{Tin-Lun Ho and C. V. Ciobanu}
\address{Department of Physics,  The Ohio State
University, Columbus, Ohio 43210}

\maketitle

\begin{abstract}
We show that the ground state of an attractive Bose gas in a double well 
evolves from a coherent state to a Schrodinger Cat like state as the 
tunneling barrier is decreased. The latter exhibits super-fragmentation 
as spin-1 Bose gas with antiferromagnetic interaction, which is caused by 
the same physics.  We also show that the fragmented condensates of 
attractive and repulsive Bose gases in double wells lead to very different 
interference patterns. 
\end{abstract}

Since the discovery of Bose-Einstein condensation (BEC) in atomic
gases, there has been many studies of the phase coherence properties 
of Bose systems. A frequently studied example is a repulsive Bose 
gas in a double well, where the ground state changes from a coherent 
state to a Fock (or number) state as tunnelling is decreased\cite{Rokhsar}
\cite{Fock}\cite{Leggett}. In a 
coherent state, the condensates in the two wells are phase coherent; whereas
they are incoherent in a Fock state. 
A number of authors\cite{JJ}\cite{Castin}\cite{Cirac}
have shown that
despite their differences, coherent states and Fock states
can not be distinguished by interference experiments which
measure the density profile of two overlapping condensates 
that are initially separated. The reason turns out to be a subtle one: 
The projective nature of the measurement process introduces coherence into a
Fock state, so much so that it gives rise to the same interference pattern
as the coherent state.
In contrast, there are little
studies of the phase coherence of Bose gases with attractive
interactions. Such studies are not only important for conceptual reasons, 
but also for current experiments\cite{Kasevich}. 

We shall see, both repulsive and attractive interactions will lead to 
``fragmented" condensates, (meaning 
more than one macroscopic eigenvalue in the single particle
density matrix). However, the ground state of an 
attractive Bose gas in double wells is in general Schrodinger-Cat like and is 
``superfragmented"\cite{HoYip}. The latter means that the internal number 
fluctuation $\Delta N^2$ is enormous, of order $\sim N^2$. In contrast, 
$\Delta N^2\sim 0$ for the Fock state (which is fragmented), and is of
order $N$ for coherent states. 
Because of this huge fluctuation, the measurement process is unable to 
project the system into a single coherent state as in the repulsive case, 
but instead a statistical mixture of Fock-like states with unequal number of 
particles in each well. 
Consequently, the interference pattern of attractive and repulsive Bose gases
in double wells are very different in the fragmented regime. 

Before proceeding, we shall comment on two recent related developments. 
There have been increasing efforts in recent years to create a
Schrodinger cat state in condensed  matter and atomic
systems.   The Schrodinger Cat state (or ``Cat state" for short)
is a superposition of two macroscopic quantum states. It is often used to
illustrate the  peculiarity of Quantum Theory, which admits such 
superpositions even though they have never been observed on the 
macroscopic scale.
The usual explanation is that Cat states 
are highly unstable against entanglement with environment\cite{ZurekPT}.
Despite such difficulties, Cat states on the mesoscopic scale
have recently been created\cite{meso}.
  Since the discovery of BEC, there
have been suggestions to produce ``bigger" Cat states using atomic Bose
condensates. These include using binary mixtures of Bose
condensates\cite{Zoller}, as well as performing projections on the coherent
states\cite{projection}.  The attractive Bose gas in a double
well is mathematically similar to the binary mixtures of Bose
condensates\cite{Zoller} but is a much simplier system. As we shall see, it
has Cat-like ground states over a wide range of parameters, a fact that 
does not seem to be generally appreciated.

In a separate development, recent studies of spin-1 Bose gases with
antiferromagnetic interaction show that strict spin conservation will 
lead to fragmented ground states, which can be Fock like (with zero spin
fluctuation) or superfragmented (with $N^2$ spin fluctuation), 
depending on the magnetization of the system\cite{HoYip}.
Although the superfragmented states of spin-1 Bose gas and attractive Bose 
gas in double wells assume very different forms, the origin of their formation
and their fluctuations are identical. 
That it is caused by the
existence of {\em degenerate minima} in
the interaction energy in number space and the quantum fluctuations between
them. In contrast, a Fock state is caused by a single deep minimum
in the interaction energy which strongly suppresses number fluctuations.
This mechanism can be seen in
spin-1 Bose gas, and is best illustrated in the double well example
below.


{\bf The Two Site Boson Hubbard Model}: Consider the Hamiltonian
\begin{equation}
H = -t( a^{\dagger}b +  b^{\dagger}a )
+ \frac{U}{2}\left[ n_{a}(n_{a}-1)  + n_{b}(n_{b}-1)  \right],
\label{BH} \end{equation}
with $n_{a}+n_{b}=N$, where $a$ and $b$ destroy Bosons 
at site (i.e. well) $a$ and $b$, $n_{a}=a^{\dagger}a$, $n_{b}=b^{\dagger}b$, 
$N$ is the total number of Bosons, $t>0$ is the tunneling  
matrix element, and $U$ is
the interaction between particles which has the same sign as the s-wave
scattering length\cite{stability}. 
Eq.(\ref{BH}) is meant to describe the actual system when reduced to 
the lowest doublet ($a\pm b$). 
Although there are residual terms in the effective Hamiltonian in 
the reduction, and that $t$ and $U$ in eq.(\ref{BH}) have 
density dependence, we shall not consider such features because they 
do not affect the basic physics of the problem (i.e. the  competition
between tunneling and interaction). Instead, we shall focus on the simple
model (eq.(\ref{BH})) which deserves to be studied in its own right. 

For a system with $N$ Bosons, the Hilbert space is 
$\{ |\ell\rangle = |\frac{N}{2}+\ell,\frac{N}{2}-\ell\rangle \}$, 
$|\ell|\leq N/2$, where $|N_{a}, N_{b}\rangle$$\equiv$
$a^{\dagger N_{a}}b^{\dagger N_{b}}$$|0\rangle$$/\sqrt{N_{a}! N_{b}!}$
is the state with $N_{a}$ and $N_{b}$ Bosons at site $a$ and $b$. 
We shall take $N$ even, a choice of convenience that has no 
effects on our results. If 
$|\Psi\rangle = \sum_{\ell} \Psi_{\ell}|\ell\rangle$ is an eigenstate 
with energy $E$, eq.(\ref{BH}) implies 
\begin{equation}
E\Psi_{\ell} = - \left( t_{\ell-1} \Psi_{\ell -1} + t_{\ell} 
\Psi_{\ell +1}
\right)  +U\ell^2 \Psi_{\ell}, 
\label{Sch} \end{equation}
where $t_{\ell} = \sqrt{\frac{N}{2}(\frac{N}{2}+1)-\ell(\ell+1)}$.
It is clear that $t_{\ell}$ strongly favors large amplitudes near $\ell=0$. 
This is reflected in the non-interacting ground state, 
$|C\rangle$$=(2^{N} N!)^{-1/2}$
$(a^{\dagger}+b^{\dagger})^{N}$$|0\rangle$, which shows 
that $\Psi^{C}_{\ell}$ is a Gaussian centered at $\ell =0$ with a width
$\sigma_{c}=\sqrt{N/2}$ since
 $\Psi^{C}_{\ell}$$\approx$$\left(\frac{2}{\pi N}\right)^{1/4} e^{-\ell^2/N}$
for $N>>1$.
When $U>0$, the potential $U\ell^2$ suppresses particle
fluctuations between  $a$ and $b$, narrowing the Gaussian toward the
delta-function $\delta_{\ell=0}$, (corresponding to the Fock state 
$|N/2, N/2\rangle$)\cite{Rokhsar}\cite{Fock}. 
The ground state for general $U>0$ (referred to as the ``repulsive family")  
is 
\begin{equation}
\Psi^{(+)}_{\ell} = G_{\sigma}(\ell),  \,\,\,\,\,\,
\frac{2}{\sigma^2}=\left(\frac{1}{N^{2}}+\frac{U}{tN}\right)^{1/2}
\label{Fockfam} \end{equation}
where $G_{\sigma}(\ell)$$= e^{-\ell^2/2\sigma^2}$$/(\pi \sigma^2)^{1/4}$.
Eq.(\ref{Fockfam})
follows from the fact that in the continuum limit 
eq.(\ref{Sch}) near $\ell=0$ is the Schrodinger equation of a simple 
harmonic oscillator. Its validity can also 
be verified numerically. 

For $U<0$, the potential $U\ell^2$ favors large amplitudes
at $\ell= \pm N/2$. It competes strongly with tunneling and tends 
to split the Gaussian $\Psi^{C}_{\ell}$ at $U=0$ into two. 
These features are found in the numerical solutions of eq.(\ref{Sch}), 
which also show to a good approximation, the ground state (referred to as the
``attractive family") is 
\begin{equation}
\Psi^{(-)}_{\ell} = Q(\Psi^{L}_{\ell} + \Psi^{R}_{\ell})
\label{Catfam}\end{equation}
where $Q$ is a normalization constant,  $\Psi^{(R)}_{\ell}$$=$
$G_{\sigma}(\ell-A)$, $\Psi^{(R)}_{\ell}$$=\Psi^{(L)}_{-\ell}$. 
(See figure 1). 
In other words, the ground state is a superposition of the 
coherent/Fock-like states
$|L\rangle$ and $|R\rangle$, 
$|\Psi^{(-)}\rangle = Q(|L\rangle + |R\rangle)$, 
\begin{equation}
|L\rangle = \sum_{q} G_{\sigma}(q)|\tilde{q}\rangle_{N_{+},N_{-}}, 
\,\,\,\,\,\, 
|R\rangle = \sum_{q} G_{\sigma}(q)|\tilde{q}\rangle_{N_{-},N_{+}},
\label{LR} \end{equation}
where $|\tilde{q} \rangle_{N_{+},N_{-}} \equiv  |N_{+}+q, N_{-}-q\rangle$,
$N_{\pm}= \frac{N}{2}\pm A$, and $N_{+}$ is the average 
number of $a$ (or $b$) particle in $|L\rangle$ (or $|R\rangle$).
We also found numerically that the quantities
$A^{\ast}=A/(N/2)$ and $\sigma^{\ast}= \sigma/\sigma_{c}$ 
$(\sigma_{c}=\sqrt{N/2})$ are functions of the combination $UN/t$
only\cite{comment}, 
with a dividing behavior at 
$UN/t=-2$ as  shown in figure 2. As $UN/t$ increases below $-2$, we have 
$A\rightarrow N/2$ and $\sigma\rightarrow 0$. The system is driven towards
the ``extreme" Cat state $|{\rm Cat}^{\ast}\rangle$
$=(|N,0\rangle+|0, N\rangle)/\sqrt{2}$. 
In fact, the overlap between $|L\rangle$ and $|R\rangle$ vanishes rapidly
when  $A>\sigma$, or
$A^{\ast}/\sigma^{\ast}>\sqrt{2/N}$. For a system with $N=1000$ particles, 
we find that $|L\rangle$ and $|R\rangle$ cease to overlap when $UN/t < -2.1$.
The system is essentially a superposition to two non overlaping mesoscopic 
condensates. 

{\bf Fragmentation and Super-fragmentation:} That interaction will lead to 
fragmentation can be seen from that
the single particle density matrices of both $|\Psi^{(+)}\rangle$ and 
$|\Psi^{(-)}\rangle$,  $\hat{\rho}$$\equiv$
$\left(\begin{array}{cc} \langle
a^{\dagger}a\rangle  & 
\langle a^{\dagger}b\rangle \\ \langle b^{\dagger}a\rangle &
\langle b^{\dagger}b\rangle \end{array}\right)$
$=\frac{N}{2}$$\left(\begin{array}{cc} 1 & x \\ x & 1 \end{array}\right)$, 
where $x= e^{-1/(4\sigma^2)}$ for $|\Psi^{(+)}\rangle$ with
$\sigma$ given in eq.(\ref{Fockfam}), and 
$x=(e^{-1/(4\sigma^2)}\sqrt{1 - (\frac{2A}{N})^2}  + e^{-(A+1/2)^2/\sigma^2})$
$/(1+e^{-A^2/\sigma^2})$. 
for $|\Psi^{(-)}\rangle$ with $\sigma$ given in fig.2. 
The eigenvalues of $\hat{\rho}$ are $\lambda_{\pm}=(N/2)(1\pm x)$. 
When $U\rightarrow 0$, we have $x=1$ for both signs of $U$, 
and $\hat{\rho}$ has only one macroscopic
eigenvalue ($\lambda_{+}=N$ and
$\lambda_{-}=0)$. This is expected since the ground state reduces to the 
coherent state $|C\rangle$. As $|U|$ increases, $x\rightarrow 0$ for both signs 
of $U$. The condensate becomes fragmented since both eigenvalues of 
$\hat{\rho}$ becomes  macroscopic, $\lambda_{+}, \lambda_{-}\rightarrow N/2$. 

However, when examining the internal number 
fluctuations $\Delta N_{a}^2$
$\equiv$$\langle(N_{a}-\langle N_{a}\rangle)^2\rangle$, 
one finds $\Delta N_{a}^2= \sigma^2/2$ for $|\Psi^{(+)}\rangle$,  
which vanishes as $U/tN$ increases. 
In contrast, 
$\Delta N_{a}^2$$=\frac{\sigma^2}{4}$$+\frac{A^2}{1+ e^{-2A^2/\sigma^2}}$
for $|\Psi^{(-)}\rangle$ which is of order $N^2$ since
$A\sim N$ and $\sigma \sim \sqrt{N}$ for large $|U|N/t$. 
The results of $\Delta N_{a}^2$  obtained from
numerical solution of eq.(\ref{Sch}) are shown in figure 2. 
We have plotted ${\rm ln}(4\Delta N^2_{a})/{\rm ln}N$ 
instead of $\Delta N_{a}^2$  because the former assumes the simple
value of 2 and 1 for the ``extreme" Cat state $|{\rm Cat}^{\ast}\rangle$
and the coherent state $|C\rangle$ respectively.  
Fig.2 shows that $\Delta N_{a}^2$ reaches a substantial fraction of 
its maximum value $N^2/4$ over the interval 
$\Delta(UN/t)$$=(-3, -2)$. Thus, from 
the viewpoint of achieving a superfragmented structure,  it is not 
necessary to go deeply into the Cat regime, (i.e. $|UN/t|>>1$).

At first sight, such an interval may seem physically irrelevant because 
for any finite interval $\Delta(UN/t)$, the corresponding range 
in $U/t$ vanishes as $1/N$ in the thermodynamic limit, and that the system 
is either deep in the Cat regime or is a coherent state depending on whether
$U<0$ or $U=0$. This, however, is not true. 
The reason is that in general, the tunneling matrix
element $t$ depends on an energy barrier $V_{o}$ in an exponential fashion,
${\rm ln} t \propto  - V_{o}$; (and $V_{o}$ is proportional to the
intensity of the Laser producing the barrier).  Thus, the range 
$\Delta V_{o}$ corresponding to 
the interval $\Delta(UN/t)$ is only proportional to ${\rm ln}N$, 
making the system highly tunable from one regime to another. 
Two other facts also make this transition region relevant. 
Firstly, quantum gases are mesoscopic 
instead of macroscopic systems, with $N<10^5$ instead 
of $N\sim 10^{23}$. Secondly, recent experiments have shown 
that the scattering  length of Rb$^{85}$ can be tuned to zero by varying 
the external magnetic field\cite{zeroasc}. 
These show that it is possible
to have a system with a small $UN$ even for $N$ as high as $10^5$.  
The fact that the ground state changes
continuously from a coherent to a Cat-like structure over a wide range of
parameter allows us to explore how the phase coherence of the system 
as the crossover takes place. 

{\bf Interference of Attractive Bose Gas in the Superfragmented regime}: 
As mentioned earlier, many authors have pointed out that there will be no
distinctions between the interference patterns of a coherent state and a Fock
state. The numerical evidence of this effect was given Javanainen 
and Yoo (JY)\cite{JJ}. Later, using analogies with quantum optics, 
Castin and Dalibard  (CD)
simulate the particle collection process by the ``beam splitter" operators 
$a\pm b$ and showed explicitly how the measurement process changes a
Fock state into a coherent state. The exact spatial pattern, however, was not
derived. In the following, we shall modify the calculation of CD to obtain the
spatial interference pattern of the attractive family eq.(\ref{Catfam}). 
We shall see that the operators for particle collection are different from the 
``beam splitter" operators, and that our calculation when applied
to the Fock state furnishes a derivation of JY's result\cite{JJ}. 

To illustrate the key features, it is sufficient to
consider the one dimensional case. At time $t=0$, the trap is turned off
and the condensates at $a$ and $b$ begins to expand and to overlap. 
For simplicity, we shall assume the atoms
expand as non-interacting particles for $t>0$, which implies 
$\hat{\psi}(x,t)= \sqrt{ \frac{m}{i 2\pi \hbar t} }$$\int {\rm d}s$
$e^{iM(x-s)^2/2\hbar t}$$ \hat{\psi}(s)$. If $a$ 
and $b$ are Wannier states localized at $\pm x_{o}$, we then have 
$\psi(s)$$\approx$$\gamma\left(e^{i\zeta (x)/2}a+e^{-i\zeta (x)/2}b\right)$, 
$\zeta(x) = (2x_{o} M/\hbar t)x $, 
and $\gamma = \sqrt{ \frac{m}{i 2\pi \hbar t} }$
$e^{iM(x^2 + x_{o}^2)/2\hbar t}$. 
If the Bosons were photons, the operators $(a+b)^{k_{+}}(a-b)^{k_{-}}$ 
represent detecting different number of photons in different beam splitters. 
However, in an interference experiment, particle detections are products of 
$\hat{\psi}(x,t)$, which are  specific combinations of $a$ and $b$ rather than
than products of $(a+b)^{k_{+}}(a-b)^{k_{-}}$.

Next we consider a series of particle detectors located at 
$x_{i}, i=1$, to $D$.
The joint probability of detecting a total of $k$ particles $(k<<N)$ 
with $k_{i}$ particles in the detector at $x_{i}$ is, (see Appendix), 
\begin{equation}
{\cal P}(\{ k_{i} \}) = 
\frac{(N-k)!}{N!} \frac{k!}{\prod_{i=1}^{D} k_{i}!}
||\hat{O}|\Psi\rangle_{N}||^2, 
\label{prob} \end{equation}
where $\hat{O}= \prod_{i} \hat{\psi}^{k_{i}}(x_{i})$ removes $k_{i}$ 
particles at $x_{i}$,  $|\Psi\rangle_{N}$ is a normalized state
with $N$ particles, and $\sum_{i=1}^{D}k_{i}=k$. 
The measured density at $x_{i}$ is given by the most probable 
set $\{ \overline{k}_{i} \}$
which optimizes ${\cal P}(\{ k_{i} \})$, i.e. $n(x_{j})=\overline{k}_{j}$. 
In case there are many such sets, the measured density will change from
experiment to experiment, as each experiment samples a different optimal set. 

For Cat like states (eq.(\ref{Catfam})), we have ${\cal P}(\{ k_{i}
\})$$=Q^2[{\cal P}_{L}(\{ k_{i} \})$
$+{\cal P}_{R}(\{ k_{i} \})$$+{\cal P}_{LR}(\{ k_{i} \})]$, where 
${\cal P}_{L}$ and ${\cal P}_{R}$ are eq.(\ref{prob}) evaluated at 
$|\Psi\rangle=|L\rangle$ and $|R\rangle$, and ${\cal P}_{LR}$ 
is eq.(\ref{prob}) with the norm replaced by 
$\langle L|\hat{O}^{\dagger}\hat{O}|R\rangle+ c.c.$. 
Since ${\cal P}_{LR}$ depends on the overlap of $|L\rangle$ and $|R\rangle$, 
it is non-vanishing only within the range of $UN/t$ such that $A<\sigma$, 
For $A>\sigma$,
${\cal P}$ is dominated  by ${\cal P}_{L}$ and ${\cal P}_{R}$, and
the interference pattern $n(x_{j})$ is proportional to
$\overline{k}^{L}_{j}$$+\overline{k}^{R}_{j}$, where 
$\{ \overline{k}^{L}_{i} \}$ and 
$\{ \overline{k}^{R}_{i} \}$ are the optimal set of ${\cal P}_{L}$ and 
${\cal P}_{R}$ respectively. 

To calculate $\{ \overline{k}^{L}_{j} \}$, we consider the coherent state
\begin{equation}
|\alpha, \beta \rangle_{_N} = \frac{1}{\sqrt{N!}}
\left(u  a^{\dagger} + v b^{\dagger}\right)^{N} |0\rangle
\label{ab} \end{equation}
where $u\equiv e^{-i\alpha/2}{\rm cos}\frac{\beta}{2}$, 
$v = e^{i\alpha/2}{\rm sin}\frac{\beta}{2}$, 
${\rm cos}^{2}\frac{\beta}{2} \equiv N_{+}/N$. 
It follows from eq.(\ref{ab}) that 
$\langle a^{\dagger}a\rangle=N_{+}$, $\langle
b^{\dagger}b\rangle=N_{-}$, $N_{\pm} \equiv N/2\pm A$. 
Eq.(\ref{ab}) has the expansion 
\begin{equation}
|\alpha, \beta\rangle_{N} 
=\sum_{q}\Psi^{(o)}_{q} e^{-i(A+q)\alpha}
|\tilde{q}\rangle_{_{N_{+}, N_{-}}}
\label{q} \end{equation}
where $\Psi^{(o)}_{q}$$=G_{\sigma_{o}}(q)$ with $\sigma_{o}^2 =
2N_{+}N_{-}/N$ up to $1/N_{+}$, $1/N_{-}$ corrections. 
Inverting eq.(\ref{q}), we can express $|\tilde{q}\rangle_{_{N_{+}, N_{-}}}$
and hence $|L\rangle$ [eq.(\ref{LR})] in terms of 
$|\alpha, \beta\rangle_{N}$. We then have 
\begin{equation}
\hat{O}|L\rangle = \Gamma \sum_{q} f(q) \int^{\pi}_{-\pi} 
\frac{{\rm d}\alpha}{2\pi} e^{i\alpha (A+q)}
\prod_{i=1}^{D}
\left[W_{x_{i}}(\alpha)\right]^{k_{i}} |\alpha, \beta\rangle_{N-k}
\label{OL} \end{equation}
where $f(q) = G_{\sigma}(q)/G_{\sigma_{o}}(q)$, 
$W_{x}(\alpha)$$=e^{i\zeta(x)/2}u + e^{-i\zeta(x)/2} v$, and 
$\Gamma = \gamma^{k}\sqrt{\frac{N!}{(N-k)!}}$. 
Since $|\alpha, \beta\rangle_{N-k}$ is close to the Fock state 
$e^{iA\alpha}$$|N_{+}, N_{-}\rangle$, 
and since the range of $q$ is restricted to $1/\sigma$, the
combination $e^{i\alpha (A+q)}$$|\alpha, \beta\rangle_{N-k}$ 
varies slowly with $\alpha$. Since $k_{i}>>1$, one can use the method of
steepest descents to determine the phase angle $\alpha^{\ast}$ 
that maximizes the magnitude $\prod_{i=1}^{D}$
$\left|W_{x_{i}}(\alpha)\right|^{k_{i}}$, and we have 
$\hat{O}|L\rangle\propto |\alpha^{\ast}\rangle$. This shows that the 
measurement process (specified by the set $\{ k_{i}\}$) projects the state
$|L\rangle$ into the coherent state $|\alpha^{\ast}, \beta\rangle$. 

To calculate ${\cal P}_{L}$, we note that 
$\langle \alpha', \beta|\alpha, \beta\rangle$
$\approx$ exp$[-(\frac{N-k}{8}{\rm sin}^{2}\beta )(\alpha'-\alpha)^2$
$+i\frac{A}{N}(N-k)(\alpha'-\alpha)]$ for $N-k>>1$. 
We then have 
\begin{equation}
{\cal P}_{L}(\{ k_{i}\}) = \eta k!
\int \frac{{\rm d}\alpha}{2\pi}
|\tilde{f}(\alpha)|^2 
\prod_{i=1}^{D}\frac{\left|W_{x_{i}}(\alpha)\right|^{2k_{i}}}{k_{i}!} 
\label{finalprob} \end{equation}
where 
$\tilde{f}(\alpha)$$=\sum_{q}f(q)$$e^{i\alpha q}$
$=\sqrt{\sigma_{o}/\sigma}$
$\sum_{q}$$e^{-q^{2}/2\sigma_{eff}^2}$$e^{i\alpha q}$, where 
$\sigma_{eff}^{-2}$$=\sigma_{o}^{-2}$$+\sigma^{-2}$, and $\eta 
=\sqrt{\frac{8\pi}{(N-k){\rm sin}^2\beta}}$.
As the ground state becomes more Cat-like, $\sigma\rightarrow 0$, 
$\tilde{f}(\alpha)$ has a weak $\alpha$ dependence. To find the optimal
set $\{ \overline{k}_{i}^{L} \}$, we rewrite the $k_{i}!$ 
in eq.(\ref{finalprob}) using 
Stirling formula, and optimize the product in eq.(\ref{finalprob}) using method
of steepest descent subject to the constraint
$\sum_{i=1}^{D}k_{i}=k$. One then obtains
$\overline{k}_{i}^{L} \propto |W_{x_{i}}(\alpha^{\ast})|^2$, or
\begin{equation}
\overline{k}^{L}_{i} = \lambda \left[ 1 + {\rm sin}\beta {\rm cos}\left(
\frac{2Mx_{o}x}{\hbar t}- \alpha^{\ast}\right) \right],
\label{kbar} \end{equation}
where $\lambda$ is a constant. 
The stationary condition for $\alpha^{\ast}$ can be
derived in a straightforward manner and will not be presented here. 
Repeating the calculation for ${\cal P}_{R}$ we find 
$\overline{k}_{i}^{R} = \overline{k}_{i}^{L}$. 
The measured density $n(x_{j})$$=\overline{k}^{L}_{j}$
$+\overline{k}^{R}_{j}$ therefore consists of a uniform background and 
a sinusoidal oscillation with wavevector $\hbar^{-1} M(2x_{o}/t)$. 
The amplitude of oscillation
${\rm sin}\beta$ vanishes as the systems becomes more
Cat-like, since $\beta\rightarrow 0$ as $A \rightarrow N/2$. 
This is in contrast to the repulsive case, where the interference pattern is
independent of the barrier height which causes the system to fragment into a
Fock state.

{\bf Appendix}. Eq.(\ref{prob}) was derived in
the Appendix in \cite{Castin} using continuous quantum
measurement theory. It can also be derived from the 
following elementary considerations: 
For a complete set of states (labelled by 
$p$) with creation operator $\{ A_{p}^{\dagger}\}$, the probability 
of detecting a 
particle in state $i$ in an $N$-particle system $|\Psi\rangle$ is 
${\cal P}^{(1)}$$= \frac{\langle 
A_{i}^{\dagger}A_{i}\rangle_{\Psi}}{N\langle 1 \rangle_{\Psi}}$
$=||A_{i}\Psi\rangle||^2/(N||\Psi\rangle ||^2)$. 
The probability of 
detecting a second particle in state $i$ after the first particle is 
detected (provided that the state evolves very little between detections) is 
${\cal P}^{(2)}$$=||A_{i}(A_{i}|\Psi\rangle)||^2$
$/[(N-1)||A_{i}|\Psi\rangle||^2]$.
The joint probability of detecting $k_{i}$ particle 
in state $i$ is ${\cal P}_{i}^{(k)}$
${\cal P}_{i}^{(k-1)}$$..{\cal P}_{i}^{(1)}$
$=\frac{(N-k)!}{N!}||A^{k_{i}}_{i}|\Psi\rangle||^2$
$/||\Psi\rangle||^2$. 
Eq.(\ref{prob}) is obtained by generalizing to the detection to more than 
one states and with $A_{i}$ replaced by $\hat{\psi}(x_{i})$. The combinatoric
factors in $k_{i}$ accounts for the arbitrary ordering in the detection of 
different $i$ states.

TLH would like to thank Mark Kasevich for calling his attention to the
similarity between attractive Bose gas in double well and the 
discussions in ref.\cite{Zoller} during the Aspen Workshop in Summer 1999, 
and for discussions of his experiments.
 This work is supported
by the NASA Grant NAG8-1441, and by the NSF Grants DMR-9705295 and DMR-9807284.

\noindent {\bf Caption}

\noindent Figure 1. Numerical solution of 
$\Psi^{(-)}_{\ell}$ calculated from eq.(\ref{BH}) for 
different $UN/t$ for a system with $N=1000$ particles. The results 
for $U<0$ can be well
fitted by the functional form in eq.(\ref{Catfam}) with parameters $A$ and
$\sigma$ shown in figure 2. 
For  $UN/t<-2$, $\Psi^{(-)}_{\ell}$ begins to split up
into two Gaussians. The split-up is complete for $UN/t\approx -2.1$.

\noindent Figure 2. With the solutions of eq.(\ref{BH}) for $U<0$ 
well fitted by eq.(\ref{Catfam}), we find that for different 
($N, U$, and $t$), each of the parameters $A^{\ast}= A/(N/2)$, 
$\sigma/\sigma_{c}$, and ${\rm ln}(4\Delta N_{a}^2)/{\rm ln} N$ 
when plotted against $UN/t$ falls into a single curve. 
Note that ${\rm ln}(4\Delta N_{a}^2)/{\rm ln} N$$=2$ and $1$ for the Cat state 
$(|N,0\rangle + |0,N\rangle)/\sqrt{2}$ and the coherent state $|C\rangle$
respectively. 
At $UN/t=-2$, ${\rm ln}(4\Delta N_{a}^2)/{\rm ln} N= 1.32$. 


\begin{references}
\bibitem{stability} For attractive Bose gas, we shall consider the density
range where the condensate in each well remains stable.
\bibitem{comment} For $U<0$, the change in potential energy and tunneling 
energy by removing a particle from $a$ to $b$ is $\sim UN$ and $t$ 
respectively, hence
the scaling variable $|U|N/t$. For $U>0$, the corresponding quantities are 
$U$ and $Nt$, and hence a different scaling parameter $U/tN$ as seen in
eq.(\ref{Fockfam}).
\end{references}
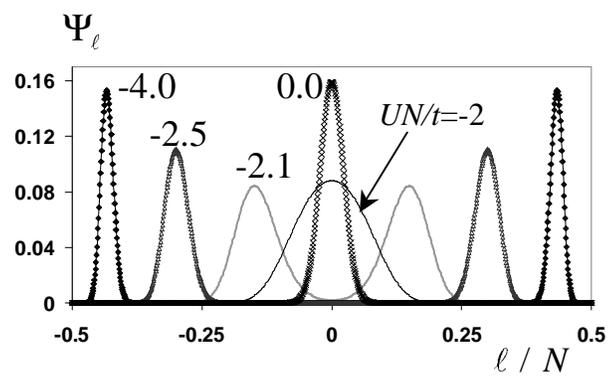
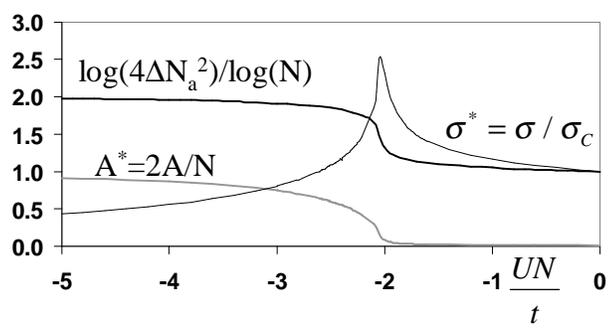
\end{document}